\documentclass[aps,prd,superscriptaddress,11pt,showpacs,notitlepage]{revtex4}
	\usepackage{graphicx}
	\usepackage{amsmath,amssymb,mathrsfs}
	\usepackage[colorlinks=true, a4paper=true, pdfstartview=FitV,
linkcolor=blue, citecolor=blue, urlcolor=blue]{hyperref}

\usepackage{epsf}
\usepackage{epsfig}
\usepackage[usenames,dvipsnames]{xcolor}

\newcommand{\be}{\begin{equation}}
\newcommand{\ee}{\end{equation}}
\newcommand{\bea}{\begin{eqnarray}}
\newcommand{\eea}{\end{eqnarray}}

\newcommand{\bb}{\bibitem}
 
\newcommand{\eqn}{\begin{eqnarray}}
\newcommand{\eqnx}{\end{eqnarray}}

\numberwithin{equation}{section}

\begin{document}

\title{The BPS property and its breaking in 1+1 dimensions}

\author{C. Adam}
\affiliation{Departamento de F\'isica de Part\'iculas, Universidad de Santiago de Compostela and Instituto Galego de F\'isica de Altas Enerxias (IGFAE) E-15782 Santiago de Compostela, Spain}
\author{A. Wereszczynski}
\affiliation{Institute of Physics,  Jagiellonian University,Lojasiewicza 11, Krak\'{o}w, Poland}

\begin{abstract}
We show that the BPS property is a generic feature of all models in (1+1) dimensions which does not put any restriction on the action. 
Here, by BPS solutions we understand static solutions which i) obey a lower-order Bogomolny-type equation in addition to the Euler-Lagrange equation, ii) have an energy which only depends on a topological charge and the global properties of the fields, but not on the local behaviour (coordinate dependence) of the solution, and iii) have zero pressure density. 

Concretely, to accomplish this program we study the existence of BPS solutions in field theories where the action functional (or energy functional) depends on higher than first derivatives of the fields. We find that  that the existence 
of BPS solutions is a rather generic property of these higher-derivative scalar field theories. Hence, the BPS property in 1+1 dimensions can be extended not only to an arbitrary number of scalar fields and k-deformed models, but also to any (well behaved) higher derivative theory.  

We also investigate the possibility to destroy the BPS property by adding an impurity which breaks the translational symmetry. Further, we find that there is a particular impurity-field coupling which still preserves {\it one-half} of the BPS-ness. An example of such a BPS kink-impurity bound state is provided. 
\end{abstract}

\maketitle 
\section{Introduction}

BPS type theories play a prominent role in current theoretical physics, where they offer an analytical insight into the nontrivial mathematical structure of topological solitons \cite{MS}, \cite{Y}. As the most famous examples, let us mention instantons in (4+0) dimensional Yang-Mills theory, the t'Hooft-Polykov monopole and the self-dual vortices in the Abelian Higgs model. A common feature sheared by {\it all} known BPS models is the property that a pertinent topological lower bound on the energy functional is saturated by solutions of lower-order, so-called Bogomolnyi equations \cite{BPS1}, \cite{BPS2}, corresponding to a first integration of the full field equations. 
Then, a linear dependence of the static energy on the pertinent topological charge is obtained, which, among other phenomena, leads to a vast moduli space and and the appearance of zero classical  binding energies. 

A simple prototype of such theories is a real scalar model in $(1+1)$ dimensions with a two-vacua potential $U$. Here, the notion of being a BPS theory refers to the following (related) features \cite{def}:
\begin{enumerate}
\item the existence of a topological lower bound on the energy which is saturated by solutions of a first order Bogomolnyi equation; 
\item the energy in the BPS sector is completely fixed by the boundary conditions on the field, i.e., fixed by the topological content of a solution rather that a particular local dependence; 
\item the Bogomolnyi equation is equivalent to the zero pressure condition.
\end{enumerate}
Note that, since the Bogomolnyi equation only applies to the static sector of the model, all this remains unchanged in the case of any type of time dynamics dictated by a suitable, physically motivated mechanism. Therefore, from now on we will refer to models in $(1+0)$ dimensions, underlining the purely static nature of our BPS solutions.  

Recently, it has been proved that these properties are preserved if one introduces an arbitrary target space $\mathcal{N}$ (i.e., considers more scalar fields) as well as if one allows for a generalised derivative term (a function of first derivatives), i.e., the so-called k-field theories \cite{FOEL}. In other words, an arbitrary model based on any number of real scalar fields $\phi_a$ (where $a=1..N$) defining a target space metric $G_{ab}(\phi)$ and involving only first spatial derivatives possesses a BPS sector with solutions saturating the corresponding Bogomolnyi equations. These formal BPS solutions are topological solitons (e.g., kinks) if the vacuum structure of the model allows for field configurations interpolating between different vacua. 

Here, the natural question arises whether scalar models with a more general energy functional, i.e., containing {\it higher derivatives}, can be of the BPS-type with solitons following from Bogomolnyi equations. The analysis of the existence of the BPS property in the most general scalar models in (1+0) dimension is the first aim of the present paper. The answer is affirmative, and {\it all} such (well-behaved) theories with arbitrary energy functional are of the BPS type (that is, allow at least for formal BPS solutions). This means that the BPS property is a {\it generic} feature of (1+1) dimensional models and, in contrast to higher dimensional cases, does not restrict the action. A second question to ask, then,  is whether it is at all possible to construct a one-dimensional scalar model which is {\it not} BPS. The investigation of the breaking of the BPS-ness is the second aim of our work. 

While our investigation is mainly of a conceptual and theoretical nature, we want to emphasize that the models considered here may perfectly well have relevant physical applications. 
Lorentz-invariant versions of higher derivative models imply higher time derivatives, and higher time derivatives, in general, lead to the well-known Ostrogradski instability, \cite{Ostro}. Nevertheless, low-energy effective field theories (LE-EFT) are usually constructed by a derivative expansion and, thus, inevitably lead to higher derivatives. Soliton solutions of these LE-EFT certainly will be relevant to understand the non-perturbative behaviour of such theories.    
On the other hand, if a higher-derivative static energy functional is interpreted as the static limit of a Lorentz non-invariant field theory (e.g., with a standard kinetic term), then this theory can, e.g., be viewed as the continuum limit of a one-dimensional discrete system (like a spin chain), with more complicated interactions between neighbouring degrees of freedom (e.g., spins), like next-to-nearest neighbours in addition to nearest neighbours.

\section{First derivative scalar theory}
We begin with a very brief summary of the results on the BPS property of scalar models in (1+0) dimensions with the standard derivative term
 \be
 E=\int_{-\infty}^\infty dx \left[ \frac{1}{2} \phi_{x}^2 +U(\phi) \right] 
 \ee
 where the potential $U$ has two isolated vacua $\phi_+^v>\phi_-^v$. 
Then, the topological lower bound is
 \bea
 E&=&\int_{-\infty}^\infty dx \left( \frac{1}{\sqrt{2}} \phi_{x} \pm \sqrt{U} \right)^2 \mp  \sqrt{2} \int_{-\infty}^\infty dx \sqrt{U} \phi_x \geq \sqrt{2} \left|  \int_{-\infty}^\infty dx \sqrt{U} \phi_x  \right| \\
 &=& \sqrt{2} \left|  \int_{\phi (-\infty)}^{\phi (+\infty)} d\phi \sqrt{U}  \right| = \sqrt{2} (\phi_+^v-\phi_-^v)^2 \left\langle \sqrt{U} \right\rangle |Q|
 \eea
where
\be
\left\langle \sqrt{U} \right\rangle =  \frac{1}{\phi_+^v-\phi_-^v} \int_{\phi_-^v}^{\phi_+^v} d\phi \sqrt{U} 
\ee
is the target space average of the potential and
\be
Q=\frac{\phi(+\infty) - \phi(-\infty)}{\phi_+^v-\phi_-^v} 
\ee
is the topological charge. For finite energy configurations $\phi(\pm \infty) \in (\phi^v_-, \phi^v_+)$ and, therefore, $Q\in  (0,\pm 1)$. The bound is saturated if and only if a first order equation (Bogomolnyi equation) is satisfied 
\be
 \frac{1}{\sqrt{2}} \phi_{x} \pm \sqrt{U}=0.
\ee
This equation is a first integral of the full static field equations (with an integration constant set to zero which guarantees finiteness of the energy). Equivalently, the Bogomonyi equation is the (local) zero pressure condition. Namely, the pressure density for the static configurations is
\be
T^{11}= \frac{1}{2} \phi_{x}^2 - U.
\ee 
These properties remain unchanged if a generalized first derivative term is added (so called k-models) as well as if a more complicated target space $\mathcal{N}$ is introduced (an arbitrary number of scalar fields) - see \cite{FOEL}, \cite{BPS-k}, \cite{Zakr1}. 

\section{Higher derivative models}
\subsection{Bogomolnyi equation and prepotential}
To analyze the existence of BPS theories in models with higher spatial derivatives, 
we consider an energy functional where the usual first derivative term is replaced by the second derivative 
 \be
 E=\int_{-\infty}^\infty dx \left[ \frac{1}{2} \phi_{xx}^2 +U(\phi) \right] . \label{energy}
 \ee
The potential is again assumed to possess two isolated vacua at $\phi^v_+ > \phi^v_-$. 
This is the simplest generalisation one may investigate, however, the obtained results are generic for any well-behaved one dimensional theory with an arbitrary number of higher-order derivatives. Some other particular examples can be found in \cite{higher-der}, \cite{higher-der-prepot} where models with $\partial_{xx}$ operator have been analyzed. Moreover, as we investigate only the static sector, the time dynamics is completely neglected. In other words, our results concern Lorentz invariant models as well as a time dependence which breaks the Lorentz invariance (for example Lifshitz theories or condensed matter systems).

Any critical points must obey 
 \be
 \delta E = \int_{-\infty}^\infty dx \left[ \phi_{xx} \frac{d^2}{dx^2} \delta \phi + U_\phi \delta \phi \right] =
 \int_{-\infty}^\infty dx \delta \phi \left[ \phi_{xxxx}  + U_\phi  \right] =0,
 \ee
 that is, 
 \be
 \phi_{xxxx}  + U_\phi =0. 
 \ee
 Let us now assume that the model has a Bogomolny type equation
 \be
 \phi_x=W(\phi) \label{BOG}
 \ee
 where $W(\phi)$ is a prepotential whose form remains to be defined. This is a direct application of the prepotential method known from the usual first-order derivative model. We remark that such a prepotential has been used also in the context of higher derivative models \cite{higher-der-prepot}. We find 
 \bea
 \phi_{xx}&=&W_x=W_\phi W \\
 \phi_{xxx} &=& \frac{d}{dx} \left(W_\phi W \right) = W_{\phi \phi} W^2 + W_\phi^2 W \\
 \phi_{xxxx} &=& \frac{d}{dx} \left(  W_{\phi \phi} W^2 + W_\phi^2 W  \right) \nonumber \\ &=& W_{\phi \phi \phi} W^3 + 4 W_{\phi \phi} W_\phi W^2 + W_\phi^3 W
 \eea
 which are just the prolongations of the first-order differential equation (\ref{BOG}).
 Hence, the prepotential equation reads
 \be
 W_{\phi \phi \phi} W^3 + 4 W_{\phi \phi} W_\phi W^2 + W_\phi^3 W +U_\phi=0.
 \ee
It is convenient to introduce a new target space function $G=W^2/2$. Then, 
\be
\phi_{xx}=G_\phi, \;\; \phi_{xxx}=G_{\phi \phi} W, \;\; \phi_{xxxx}= 2 G_{\phi \phi \phi} G+G_{\phi \phi} G_\phi
\ee
and
\be
2 G_{\phi \phi \phi} G+G_{\phi \phi} G_\phi + U_\phi=0
\ee
Hence,

\be
2 \frac{d}{d\phi} (G_{\phi \phi} G) -  \frac{1}{2} \frac{d}{d\phi} G_{\phi}^2  + U_\phi=0
\ee
which can be integrated to
\be
2G_{\phi \phi} G -\frac{1}{2} G_\phi^2 + U = C. \label{pre}
\ee
Below we show that the integration constant should be chosen $C=0$. 
Formulas (\ref{BOG}), (\ref{pre}) with $C=0$, define the pertinent Bogomolny equations or, in other words, the BPS sector of the model. Note that the higher derivative dependence of the energy functional is reflected in a more involved form of the prepotential equation, which now is a second-order nonlinear ordinary differential equation on target space 
for which finding a solution may be rather difficult. Therefore, constructing exact solitonic solutions for higher derivative models, even in the BPS sector, is a more complicated task than in the usual first derivative theories. 

Although we considered a specific example, we want to underline the general character of the procedure. Indeed, after assuming $\phi_x = W(\phi)$ for an arbitrary energy density $\mathcal{E}(\phi ,\phi_x ,\phi_{xx})$, we may insert this relation and its prolongations into the Euler-Lagrange (EL) equation for $\mathcal{E}$, converting the fourth-order EL equation into a third-order target space equation for $W$ (the derivative of the prepotential equation). Further, for translationally invariant energy functionals, the EL equation always has a "trivial" first integral, corresponding to momentum conservation (or constant pressure). For our concrete example, this first integral reads
\be \label{first-int}
\phi_{xxx}\phi_x - \frac{1}{2} \phi_{xx}^2 + U = C.
\ee
But this first integral directly implies a first integral for the third-order target space equation, resulting in the second-order prepotential equation. Indeed, if we insert Eq. (\ref{BOG}) and its prolongations into the above first integral (\ref{first-int}), we directly re-derive the prepotential equation (\ref{pre}).  

There are several comments in order. First of all, a physically acceptable solution of the prepotential equation must obey the following boundary condition
\be
G_\phi (\phi^v) =0 \;\;\; \wedge \;\;\; \left( G (\phi^v) =0 \;\; \vee \;\; G_{\phi \phi} (\phi^v) =0 \right) .
\ee
The first condition follows from the finiteness of the energy, while the second is a further consequence of the prepotential equation. At the moment, we cannot prove that this equation has an acceptable solution for any positive definite double vacuum potential. Of course, another important issue is the uniqueness of such a solution.

Secondly, similarly as in the usual quadratic first derivative models, one could apply an inverse logic to the prepotential equation and construct the potential $U$ for a given prepotential $G$. However, it is not guaranteed that for $G$ obeying the above boundary conditions the derived $U$ is positive definite. In fact, it is easy to provide a counter example. Let us assume $G=(1-\phi^2)^2$ (which for the standard energy integral corresponds to the $\phi^4$ model). Then, $G_\phi=-4\phi (1-\phi^2)$, $G_{\phi\phi}=4(3\phi^2-1)$. This gives
$U=8(1-\phi^2)^2(1-2\phi^2)$. 

Thirdly, a critical point does not have to correspond to a minimum energy solution. This is clearly visible even in the case when we neglect the potential. Then, the prepotential equation possesses two exact solutions: $G=A$ and  $G=B \phi^{4/3}$. The first, constant solution gives the correct energy minimum (vacuum) with $\phi = \pm \sqrt{2A} ( x-x_0)$. However, the second solution leads to the infinite energy solution $\phi=\pm (\sqrt{2B}/3)^3 (x-x_0)^3$. Note that only the first, constant solution obeys the above boundary conditions for the prepotential. 

One has to remember that a particular choice of derivative terms in the static energy functional is reflected in the prepotential equation. As a consequence, the issue of the existence (and uniqueness) of the prepotential should be investigated for each case separately. 

As a fourth comment, let us mention that for standard energy densities $\mathcal{E}(\phi ,\phi_x)$ a rather general reduction-of-order method has been developed recently under the name of "First-Order Euler-Lagrenge" method \cite{FOEL}, which allows to recover all known BPS solutions. This method may be generalised to higher derivative theories and allows to re-derive the BPS equations and solutions of this section. As the presentation and the application of the method covers several pages, we relegate it to an appendix. 

\vspace*{0.2cm}

It can be instructive to consider a model where the higher derivative part is a small correction to the usual first order kinetic term
 \be
 E_\epsilon=\int_{-\infty}^\infty dx \left[ \frac{1}{2}\phi_x^2+  \frac{\epsilon}{2} \phi_{xx}^2 +U(\phi) \right] .
 \ee
 Such a model is physically more reasonable, as in the small field limit (close to vacuum limit) we recover the usual model quadratic in the first derivative. Thus, the higher derivative addition can be treated as an effective term relevant at higher order. 
Then, the prepotential equation is ($C=0)$
\be
-G + \epsilon \left( 2G_{\phi \phi} G -\frac{1}{2} G_\phi^2  \right)+ U = 0.
\ee
Obviously, in the limit $\epsilon=0$ we obtain the prepotential known from the standard scalar model. 
\subsection{Energy in the BPS sector}

 For solutions of the Bogomolny equation, the energy can be written as a target space integral. Namely, 
 \bea
 E&=&\int_{-\infty}^\infty dx \left[ \frac{1}{2} \phi_{xx}^2 +U(\phi) \right] = \int_{-\infty}^\infty dx \left[ \frac{1}{2} \phi_{xx}^2 +U(\phi) \right]  \frac{\phi_x}{\phi_x} \\ 
 &=& \int_{-\infty}^\infty dx \left[ \frac{1}{2} W_\phi^2 W +\frac{U(\phi)}{W} \right] \phi_x = \int_{\phi (-\infty)}^{\phi (+\infty)} d\phi \left[ \frac{1}{2} W_\phi^2 W +\frac{U(\phi)}{W} \right] \\
&=& \int_{\phi (-\infty)}^{\phi (+\infty)} \frac{d\phi}{\sqrt{2G}} \left[ \frac{1}{2} G_\phi^2 +U(\phi) \right] .
 \eea
 This means that the energy depends only on the model (is a target space average of a target space  function) and the topological charge (i.e., the boundary conditions). 
 \be
 E=(\phi^v_+-\phi^v_-)^2 \left\langle  \frac{G_\phi^2 +2U(\phi)}{2\sqrt{2G}}  \right\rangle |Q|
 \ee 
 where now
 \be
 \left\langle  \frac{G_\phi^2 +2U(\phi)}{2\sqrt{2G}}  \right\rangle \equiv   \frac{1}{\phi^v_+ - \phi^v_-} \int_{\phi_-}^{\phi_+} \frac{d\phi}{2\sqrt{2G}} \left[  G_\phi^2 +2U(\phi) \right] .
 \ee
 This is a well-known and typical behaviour of BPS solutions.

\subsection{Zero pressure condition}
Finally, we show that the Bogomolny equation is equivalent to the zero pressure condition. 
For the Lorentz invariant Lagrangian the energy momentum tensor reads 
 \be
 T^{\alpha \mu} = - \partial_\beta \frac{\partial L }{\partial \partial_{\alpha \beta} \phi} \partial^\mu \phi 
 + \frac{\partial L }{\partial \partial_{\alpha \beta} \phi}  \partial_\beta \partial^\mu \phi - \eta^{\alpha \mu}L.
 \ee
For the static sector of the above model (defined by the energy integral (\ref{energy})) we find 
 \be
 T^{11}=- \partial_x \frac{\partial \mathcal{E} }{\partial \partial_{xx} \phi} \partial^x \phi 
 + \frac{\partial \mathcal{E} }{\partial \partial_{xx} \phi}  \partial_x \partial^x \phi - \eta^{xx}\mathcal{E}=\partial_x \frac{\partial \mathcal{E} }{\partial  \phi_{xx}}  \phi_x
 - \frac{\partial \mathcal{E} }{\partial  \phi_{xx}}   \phi_{xx} + \mathcal{E}
 \ee
 where $\mathcal{E}$ is the energy density. Hence
 \be
 T^{11}= \phi_{xxx} \phi_x - \phi_{xx}^2 +\frac{1}{2} \phi_{xx}^2 + U = \phi_{xxx}\phi_x -\frac{1}{2}\phi_{xx}^2+U .
 \ee
 In the BPS sector the zero pressure condition can be written as
 \be
2G_{\phi \phi} G-\frac{1}{2} G_\phi^2+U=0 .
 \ee
 Thus, the BPS equation or, strictly speaking, the prepotential equation with $C=0$ is equivalent to the zero pressure condition. 
 
\subsection{An example}
As an exact example we consider the following energy integral 
 \be
 E_{\epsilon=1} =\int_{-\infty}^\infty dx \left[ \frac{1}{2}\phi_x^2+  \frac{1}{2} \phi_{xx}^2 +\frac{3}{2} (1-\cos \phi)^2 \right] 
 \ee
where the usual quadratic term in first derivatives is accompanied by a second derivative counterpart. Furthermore, the potential is chosen as the sine-Gordon potential squared. For this model, one can solve the prepotential equation and finds
\be
G=1-\cos \phi
\ee
which gives a kink profile identical to the standard sine-Gordon soliton 
\be
\phi(x)=4\arctan e^{\pm (x-x_0)}.
\ee
The energy in the BPS sector reads
\be
E=\int_{\phi_-^v}^{\phi^v_+} \frac{d\phi}{\sqrt{2G}} \left[ G +\frac{1}{2} G_\phi^2 +U\right] = 2 \int_{\phi_-^v}^{\phi^v_+} \frac{d\phi}{\sqrt{2G}} \left[ G +\frac{1}{2} G_\phi^2 -GG_{\phi \phi}\right]
\ee
which for the above prepotential gives $E=40/3$. 

\subsection{KdV equation}
Although this goes beyond the scope of the present paper, it is interesting to notice that the KdV equation can be partially put into the BPS framework, as well. Of course, the model does not support static topological solitons. On the other hand, there are non-topological traveling solitary waves which are solution of the following KdV equation
\be
\phi_t +\phi_{yyy}-6\phi \phi_y=0.
\ee 
The one-soliton solution can be constructed under the traveling wave assumption, i.e., $\phi(t,y)=\phi(y-ct)$. Then, we get
\be
-c\phi_x +\phi_{xxx}-6\phi \phi_x=0 \label{KdV-BPS}
\ee
where $x=y-ct$. This equation can be viewed as a 'static' higher order equation in the new variable $x$. Now, we assume that again $\phi_x=\sqrt{2G}$ which leads to the following equation for $G$
\be
-c-6\phi + G_{\phi \phi}=0.
\ee
As in the previous BPS models, this equation can be integrated at least once. Notably, due to its linearity one can find $G$ exactly (the two integration constant are assumed to vanish) 
\be
G=\frac{c}{2} \phi^2 + \phi^3.
\ee
Hence,
\be
\frac{1}{2}\phi_x^2=\frac{c}{2} \phi^2 + \phi^3 \label{KdV-BOG}
\ee
which is solved by 
\be
\phi=-\frac{c}{2} \frac{1}{\cosh^2 \frac{\sqrt{c}}{2}x }
\ee
that is, the usual one-solitary wave profile. In other words, as in all previous examples, the one-soliton configuration can be treated as a BPS solution.

Furthermore, the field equation in the BPS sector (\ref{KdV-BPS}) follows from the following pseudo-energy integral 
\be
F[\psi]= \int \left( \frac{c}{2}\psi_x^2 +\psi_x^3+\frac{1}{2} \psi_{xx}^2\right) dx .
\ee 
Indeed, the variational principle leads to $-c\psi_{xx}-3\partial_x(\psi_x^2) +\psi_{xxxx}=0$, which 
after an identification $\phi=\psi_x$ gives (\ref{KdV-BPS}). Remarkably, this identification inserted directly into the pseudo-energy 
\be
F[\phi]= \int \left( \frac{c}{2}\phi^2 +\phi^3+\frac{1}{2} \phi_{x}^2\right) dx
\ee
 immediately results in the Bogomolnyi equation (\ref{KdV-BOG}). 
\section{Breaking the BPS-ness}
\subsection{Impurity and breaking of the translational invariance}
As we saw, the BPS property is a phenomenon shared (probably) by all (multi)scalar models with any derivative term. In order to construct a model which does not possess this feature, one possibility consists in breaking the translational invariance, for example by introducing an impurity. There are many ways in which an impurity has been considered in the context of topological kinks \cite{impurity}. Here we introduce it in the simplest way where the standard scalar model with a potential is accompanied by an impurity $\sigma (x)$
 \be
 E=\int_{-\infty}^\infty dx \left[ \frac{1}{2} \phi_{x}^2 +U(\phi) + \phi \sigma(x) \right] .
 \ee
Then the static equation of motion reads
\be
\phi_{xx} - U_\phi - \sigma(x)=0 .
\ee
This expression is no longer integrable to a first order equation. Furthermore, the zero pressure condition does not imply the static field equations. Indeed, the pressure density is 
\be
T^{11}=\frac{1}{2} \phi_{x}^2 -U(\phi) - \phi \sigma(x)
\ee
and leads to 
\be
\phi_x \left( \phi_{xx} -U_\phi -\sigma(x) \right) = \phi \sigma_x
\ee
which together with the static equation of motion gives $\phi \sigma_x =0$. This is in contradiction with the fact that the introduced impurity is nontrivial (not a constant). In other words, solutions of the model do not correspond to zero pressure density configurations, which is sufficient to show that the model does not possess the BPS property. 

A solvable example of such a model can be given if the impurity has functionally the same form as the second derivative of the scalar field in the model without the impurity. As an example, we can consider the sine-Gordon model with the following impurity
 \be
 E=\int_{-\infty}^\infty dx \left[ \frac{1}{2} \phi_{x}^2 + (1-\cos \phi) + \phi \sigma(x) \right] 
 \ee
where
\be
\sigma(x)=\frac{2  \alpha}{1-\alpha} \frac{\sinh \frac{ x}{\sqrt{1-\alpha}} }{\cosh^2 \frac{ x}{\sqrt{1-\alpha}} }
\ee
and $\alpha<1$ is a real parameter of the impurity which modifies its width. Note that such a form of the impurity is physically well motivated as it is exponentially localised. 
Then, the exact static anti-kink solution is 
\be
\phi = 4\arctan e^{- \frac{1}{1-\alpha} x }.
\ee
The model not only breaks the translational invariance. The usual $\mathbb{Z}$ target space symmetry $\phi \rightarrow - \phi$ is replaced by $(\phi, \sigma) \rightarrow (-\phi , -\sigma)$. Therefore, there is no obvious way to construct the charge conjugate kink solution in an exact form. However, the lack of this target space symmetry is not related to the breaking of the BPS-ness of the scalar field model. In fact, one can consider a $\mathbb{Z}_2$ invariant model which still does not support BPS solitons. Namely, we may introduce a position dependent "mass" term
\be
 E=\int_{-\infty}^\infty dx \left[ \frac{1}{2} \phi_{x}^2 +U(\phi) +\frac{m^2(x)}{2} \phi^2  \right] 
 \ee
 whose solutions again are not zero pressure configurations. 
\subsection{Impurity which preserves the BPS property}
 An interesting observation is that it is possible to break the translational invariance by adding an impurity in such a way that the BPS property is preserved. This resembles the situation in the Abelian Higgs model at critical coupling where  it is known how to introduce a translational symmetry breaking impurity which however keeps the BPS property unaffected \cite{vortices}, \cite{vortices-2} (while of course the pertinent Bogomolnyi equations are modified). The required modification of the energy functional is 
 \be
 E=\int_{-\infty}^{\infty} dx \left[\frac{1}{2} \phi_x^2 + U(\phi) + 2\sigma \sqrt{U} + \sqrt{2} \sigma \phi_x\right] +\int_{-\infty}^{\infty} dx \sigma^2 \label{BPS-imp}
\ee
where the last term can be viewed as the energy of the impurity which obviously does not change the field equations. 
 Then the topological bound is
 \bea
 E&=&\int_{-\infty}^{\infty} dx \left( \frac{1}{\sqrt{2}} \phi_x + (\sigma + \sqrt{U})\right)^2  -\sqrt{2} \int_{-\infty}^{\infty} dx \phi_x \sqrt{U} \\
 &\geq& -\sqrt{2} \int_{-\infty}^{\infty} dx \phi_x \sqrt{U} =  \pm \sqrt{2} \int_{\phi^v_-}^{\phi^v_+} d\phi \sqrt{U} 
 \eea
 where the sign depends on the asymptotical values of the scalar field, which is controlled by the strength of the impurity. 
 
 The bound is saturated if the following Bogomolnyi equation is obeyed 
 \be
  \frac{1}{\sqrt{2}} \phi_x + \sigma + \sqrt{U}=0
 \ee
 This equation implies also the full static equation of motion. Indeed, it gives
 \be
  \frac{1}{\sqrt{2}} \phi_{xx} + \sigma_x + \frac{1}{2\sqrt{U}} U_\phi \phi_x=0\;\; \Rightarrow \;\; 
 \phi_{xx} + \sqrt{2} \sigma_x - \frac{U_\phi}{\sqrt{U}} (\sigma + \sqrt{U})=0
 \ee
which is exactly the Euler-Lagrange equation for the above energy functional,
\be
\phi_{xx}-U_\phi-\sigma \frac{U_\phi}{\sqrt{U}} +\sqrt{2}\sigma_x=0.
\ee
The value of the energy of the BPS solution, or strictly speaking its sign, is decided by the strength of the impurity. In fact, for a solution of the Bogomolnyi equation the derivative of the scalar field is governed by a mutual effect of the potential and the impurity
\be
\frac{1}{\sqrt{2}} \phi_x =-\sqrt{U}-\sigma .
\ee
If the impurity is positive everywhere then $\phi_x<0$ which means that $\phi(x=-\infty)=\phi_+$ and $\phi(x=\infty)=\phi_-$. Then, the BPS kink-impurity bound state has a negative topological charge and the energy is positive $E=  \sqrt{2} (\phi_+^v-\phi_-^v)^2 \left\langle \sqrt{U} \right\rangle $. On the other hand, for a sufficiently negative impurity such that $\phi_x >0$ for all $x$ we get a BPS solution with a positive topological charge whose energy is negative and reads $E= - \sqrt{2} (\phi_+^v-\phi_-^v)^2 \left\langle \sqrt{U} \right\rangle $.

Finally, let us consider the pressure density 
\be
T^{11}=\frac{1}{2} \phi_x^2- U(\phi) - 2\sigma \sqrt{U} -\sigma^2=\frac{1}{2} \phi_x^2 - (\sigma +\sqrt{U})^2
\ee
Hence, one of the two roots of the zero pressure equation coincides with the Bogomolnyi equation. All that proves that this particular coupling of the impurity does not destroy the BPS-ness of the scalar theory, even though the translational invariance is broken. 

The fact that only one root of the zero pressure condition leads to the Bogomolnyi equation results in a kink-antikink asymmetry.  Only the kink (or antikink) is a solution of the Bogomolnyi equation saturating the topological bound. The topological charge conjugate solution does not obey the corresponding first order equation. Perhaps, one could call this BPS completion of the scalar model with an impurity a {\it  half-BPS} model. This is exactly what happens in the Abelian Higgs model at the critical coupling where the vortices remain BPS objects, again with a positive or negative value of the topological charge \cite{vortices}, \cite{vortices-2} (depending on the sign of the impurity coupling in the action). One difference is that in the Abelian Higgs model the impurity is coupled to the magnetic field while here $\sigma(x)$ couples to the topological current (the derivative of the scalar field). On the other hand, both quantities are pullbacks. 

\vspace*{0.2cm}

As an exact example, let us consider the sine-Gordon potential $U=1-\cos \phi$ with an impurity of the following form
\be
\sigma(x)=\sqrt{2}(\alpha-1) \frac{1}{\cosh \alpha x}
\ee
where $\alpha \in \mathbb{R}-\{0\}$. Then, the BPS kink-impurity bound state solution, which in other words describes the kink localised on the impurity, is 
\be
\phi_k(x)=4\arctan e^{-\alpha x}
\ee
and the energy saturates the topological bound $E=\pm 8$. Note that this solution has positive topological charge (negative energy) for $\alpha <0$ and negative charge (positive energy) for $\alpha >0$.  

\section{Summary}
We have shown that the concept of BPS theories can be extended to one-dimensional  (multi)-scalar field theories with higher derivatives. Specifically, we have found that: 
\begin{enumerate}
\item the static equations of motion admit a reduction to a first order equation with the prepotential obeying a pertinent nonlinear differential equation on the target space; 
\item the energy in the BPS sector depends only on the boundary values of the field and therefore is proportional to the topological charge(s), while the proportionality constant encodes information on the particular model via a target space average of a function of the prepotential; 
\item the prepotential equation is exactly the zero pressure condition in the static sector. 
\end{enumerate}

The particular form of the prepotential equation is closely related to the derivative terms of the model. For example, if the term $\phi_{xx}^2$ is added to the static energy functional then we arrive at a second order nonlinear ODE for the prepotential. This makes the analysis of the equation very complicated. As a consequence, the existence of a prepotential for this model (as well as for other generalised models) is a rather involved problem. However, we presented some examples for which a prepotential does exist. Thus these examples are genuine BPS theories. 

All these three properties are the usual properties defining the standard BPS sector of a scalar field theory with at most first spatial derivatives. Hence, there is no substantial difference between the usual models and higher derivative generalizations. This also suggests that the BPS feature in $(1+0)$ dimensions is a completely common property which, in contrast to higher dimensional models, does not select any particular theory. In other words, the BPS feature is to some extent a {\it trivial phenomenon} in the case of one spatial dimension. For example, it seems that {\it all} Lorentz invariant scalar models  (with any generalized derivative term) enjoy this property. This should be contrasted with BPS models in higher dimensions where the BPS property is a highly restrictive requirement which puts strong restrictions on the action. As a consequence, only very few solitonic models in higher dimensions possess the BPS property. One may propose several explanations for such a difference. First of all, in one dimension a BPS theory supports only kink or anti-kink solution while multi-kink configurations usually occur beyond the BPS sector (except compacton solutions). In higher dimensions, the BPS theory allows for {\it infinitely} many multi-soliton solutions which obviously is much more difficult to guarantee. 

On the other hand, we found one way to destroy the BPS-ness. This can be achieved if a translational symmetry breaking term is added. Whether there exists another mechanism to destroy the BPS property or it is truly intimately related to the breaking of the translational invariance remains to be further analysed. 

Rather surprisingly, we found that it is possible to couple an impurity (for instance in the usual first derivative squared model) in such a manner that the BPS property is partially maintained. It means that the kink (or antikink) obeys a Bogomolny equation and therefore saturates a pertinent topological bound. The topological charge conjugate solution, however, only obeys a full second-order field equation and possesses higher energy which exceeds the bound. Due to the observed asymmetry between the BPS kink and the non-BPS antikink, such a model may be used to investigate the role of BPS-ness in the dynamics of topological solitons in (1+1) dimensions. For that, one should compare the scattering of two kinks (BPS objects) with the scattering of two antikinks (non BPS objects).  One can also ask about supersymmetric extensions of these half-BPS models, as well as of higher derivative BPS scalar theories, applying already developed methods  \cite{susy}.

Furthermore, it is possible to construct impurity BPS models in {\it higher} dimensions. This follows from the observation that the derivative part of the impurity BPS model considered here can be written as the topological current squared $j_\mu \sim \epsilon_{\mu \nu} \partial^\nu \phi$. This current has its higher $d$ dimensional generalisation $j_\mu \sim \epsilon_{\mu \mu_1..\mu_d} \epsilon^{a_1..a_d} \partial^{\mu_1} \phi^{a_1}...\partial^{\mu_d} \phi^{a_d}$ which of course requires $d$ scalar fields (defining a $d$-dimensional target space $\mathcal{N}$). 
Then, the construction of the impurity BPS model (with, for example, a radial impurity) should follow exactly the same lines as in the one dimensional case of the present paper. This means that the $(3+1)$ dimensional BPS Skyrme model \cite{BPS Sk} as well as its baby version \cite{bBPS} should possess a BPS-ness preserving extension where an impurity is added. 

\section*{Acknowledgements}
The authors acknowledge financial support from the Ministry of Education, Culture, and Sports, Spain (Grant No. FPA2017-83814-P), the Xunta de Galicia (Grant No. INCITE09.296.035PR and Conselleria de Educacion), the Spanish Consolider-Ingenio 2010 Programme CPAN (CSD2007-00042), Maria de Maetzu Unit of Excellence MDM-2016-0692, and FEDER.

\appendix
\section{The Reduced-Order Euler-Lagrange equations}
For standard energy functionals which only depend on the fields themselves and their first derivatives,  a reduction-of-order method capable of reproducing all known BPS solutions has been introduced recently under the name of the "FOEL" (First-Order Euler-Lagrange) method \cite{FOEL}. The FOEL method, which replaces the second-order EL equations by a set of first-order "FOEL" equations, is a generalisation of the "concept of strong necessary conditions" developed e.g. in \cite{sok1}. Adapted to the one-dimensional models considered here, the FOEL method is based on the following two simple observations. Firstly, for an energy density $\mathcal{E}(\phi,\phi_x)$, the two FOEL equations
\be
\frac{\partial\mathcal{E}}{\partial \phi}=0, \qquad \frac{\partial\mathcal{E}}{\partial
{\phi_x}}=0
\ee
 are sufficient conditions for the EL equation 
 \be
 \frac{\partial\mathcal{E}}{\partial \phi} - D_x \frac{\partial\mathcal{E}}{\partial {\phi_x}}=0.
 \ee
 In this appendix we use the notation $D_x $ for the total $x$ derivative
 $D_x = \partial_x + \phi_x \partial_\phi + \phi_{xx} \partial_{\phi_x} + \ldots$ to distinguish it from the partial derivative $\partial_x$ which only acts on the explicit $x$ dependence of a function $f(x,\phi,\phi_x ,\ldots )$.  
Due to their restrictive nature, the FOEL equations usually only permit trivial solutions. The second observation is that the FOEL equations, in contrast to the EL equations, are {\em not} invariant under the addition of total derivatives to the energy density, $\mathcal{E} \to \mathcal{\bar E} = \mathcal{E} - D_x F(x, \phi) $ where $F$, in general, is an arbitrary function of its arguments. Depending on the choice of $F$, the resulting FOEL equations (and using $D_x F = F_{,x} + F_{,\phi} \phi_x$)
\bea
 \frac{\partial\mathcal{\bar E}}{\partial \phi}&=&\frac{\partial\mathcal{E}}{\partial \phi} - F_{,x\phi} + F_{,\phi\phi} \phi_x =0 \\ \nonumber
 \frac{\partial\mathcal{\bar E}}{\partial {\phi_x}} &=& \frac{\partial\mathcal{E}}{\partial {\phi_x}} - F_{,x\phi_x} - F_{,\phi\phi_x} \phi_x - F_{,\phi} =0
 \eea
 may, therefore, lead to nontrivial solutions for the {\em same} system of EL equations. In particular, all known BPS solitons, instantons, but also solutions produced by B\"acklund transformations can be recovered within the FOEL method.
 
 There is an obvious generalisation of the FOEL method for energy functionals which depend on higher derivatives. Concretely, for an energy density $\mathcal{E}(\phi,\phi_x, \phi_{xx}) $ with EL equation
 \be
 \frac{\partial\mathcal{E}}{\partial \phi} - D_x \frac{\partial\mathcal{E}}{\partial {\phi_x}} + D_xD_x \frac{\partial\mathcal{E}}{\partial \phi_{xx}}=0
 \ee
the following three equations
\be
\frac{\partial\mathcal{E}}{\partial \phi}=0, \qquad \frac{\partial\mathcal{E}}{\partial
{\phi_x}}=0, \qquad \frac{\partial\mathcal{E}}{\partial \phi_{xx}}=0
\ee
are sufficient conditions for the EL equations. These equations form a system of second order equations, in contrast to the EL equation which is of fourth order. For a general energy density which depends on field derivatives up to order $n$, the corresponding restricted equations (sufficient conditions for the EL equations) are of order $n$, whereas the EL equations are of order $2n$.
We shall, therefore, call the equations "Reduced-Order Euler-Lagrange" (ROEL) equations and the method based on them the ROEL method. Adding a total derivative, as before (and assuming $F=F(\phi ,\phi_x)$ for simplicity, i.e., no explicit $x$ dependence)
\be
\mathcal{\bar E}(\phi ,\phi_x ,\phi_{xx}) = \mathcal{E} (\phi ,\phi_x ,\phi_{xx}) - D_x F(\phi ,\phi_x) \equiv \mathcal{E} - F_{,\phi}\phi_x - F_{,\phi_x} \phi_{xx},
\ee
the resulting ROEL equations are
\bea
\frac{\partial \mathcal{\bar E}}{\partial \phi} &=& \frac{\partial \mathcal{ E}}{\partial \phi} - F_{,\phi\phi} \phi_x - F_{,\phi \phi_x} \phi_{xx} =0, \label{roel1} \\
\frac{\partial \mathcal{\bar E}}{\partial \phi_x} &=& \frac{\partial \mathcal{ E}}{\partial \phi_x}- F_{,\phi\phi_x} \phi_x - F_{,\phi} -F_{,\phi_x\phi_x}\phi_{xx} =0, \label{roel2} \\
\frac{\partial \mathcal{\bar E}}{\partial \phi_{xx}} &=&\frac{\partial \mathcal{ E}}{\partial \phi_{xx}}
- F_{,\phi_x} =0. \label{roel3}
\eea
The three equations have the common first integral $\mathcal{\bar E}\vert =0$ or
\be \label{roel4}
\mathcal{E}\vert = \left( F_{,\phi}\phi_x + F_{,\phi_x}\phi_{xx} \right) \vert
\ee
where the vertical line means that the expression is evaluated on-shell, i.e., on a ROEL solution.

We remark that, in addition to the full set of ROEL equations (\ref{roel1})-(\ref{roel3}), in the case of higher-derivative theories it is possible to use partial ROEL equations like, e.g., 
\bea
\frac{\partial \mathcal{\bar E}}{\partial \phi} &=& \frac{\partial \mathcal{ E}}{\partial \phi} - F_{,\phi\phi} \phi_x - F_{,\phi \phi_x} \phi_{xx} =0, \label{proel1} \\
- \frac{\partial \mathcal{\bar E}}{\partial \phi_x} + D_x \frac{\partial \mathcal{\bar E}}{\partial \phi_{xx}} &=& - \frac{\partial \mathcal{ E}}{\partial \phi_x} + D_x \frac{\partial \mathcal{ E}}{\partial \phi_{xx}} 
 + F_{,\phi} =0
 \eea
 which are, in general, less restrictive and may give rise to solutions which do not solve the full ROEL equations \footnote{These partial ROEL equations were already introduced in \cite{sok1} under the name of "semi-strong necessary conditions" and applied to the KdV equation.}. 
For our purpose of reproducing the results of Section III.A, both the full and the partial ROEL equations 
work, so we shall restrict to the full ROEL equations, for simplicity.  

For the simple energy density $\mathcal{E} = (1/2) \phi_{xx}^2 + U(\phi)$ of Section III.A, the third ROEL equation reads $\phi_{xx} = F_{,\phi_x}$. Now we assume that the solution which we want to find allows to express $\phi_{xx}$ as a function of $\phi$ only, $\phi_{xx} = f(\phi)$, because we know that the solution which we want to reproduce has this property. This assumption implies that $F$ is at most linear in $\phi_x$, i.e.,
\be
F = A(\phi) + B(\phi)\phi_x \quad \Rightarrow \quad \phi_{xx} = B(\phi).
\ee  
Inserting this into the second ROEL equation, we easily find
\be
\phi_x = - \frac{A_{,\phi}}{2B_{,\phi}},
\ee
that is, also $\phi_x$ can be expressed as a function of $\phi$ only, as required.
Inserting both expressions into the first integral (\ref{roel4}) we get for the potential
\be
U=\frac{1}{2}B^2 - \frac{1}{4}\frac{A_{,\phi}^2}{B_{,\phi}} .
\ee
But we still have to implement the consistency condition
\be
B = \phi_{xx} = D_x \phi_x = D_x \left( -\frac{A_{,\phi}}{2B_{,\phi}} \right).
\ee
As we found that $\phi_x$ is equal to a function of $\phi$ only, we may call this function $W$, without loss of generality,  i.e., $\phi_x = W(\phi)$. Its first prolongation is then $\phi_{xx} = WW_{,\phi}$, and we get for $A$ and $B$
\be
B = WW_{,\phi} , \quad A_{,\phi} = -2 WB_{,\phi} = -2(WW_{,\phi}^2 + W^2W_{,\phi\phi}). 
\ee
Inserting these expressions, we get for the potential
\be
U = -\frac{1}{2}W^2W_{,\phi}^2 - W^3 W_{,\phi\phi} = \frac{1}{2}G_{,\phi}^2 - 2GG_{,\phi\phi}
\ee
(where $W=\sqrt{2G}$) which exactly coincides with Eq. (\ref{pre}) for $C=0$.

So, indeed, we were able to reproduce the results of Section III.A within the ROEL formalism. The ROEL calculation is, however, more involved than the direct calculation starting from $\phi_x = W(\phi)$. The reason is that, starting from $\phi_x = W$, its prolongations (higher derivatives of $\phi$) can be calculated by direct differentiation, which is simple. The ROEL method, on the other hand, requires to start from the highest derivative (here $\phi_{xx} = B$), and lower derivatives have to be found by a kind of integration, which is usually more complicated. 
The ROEL method may, however, also be useful to find more general, non-BPS  solutions for these higher-derivative theories, in particular, when time dependence is taken into account.


\end{document}